\documentclass[a4paper,fleqn,usenatbib]{mnras}

\usepackage{newtxtext,newtxmath}

\usepackage[T1]{fontenc}
\usepackage{ae,aecompl}

\usepackage{graphicx}	
\usepackage{amsmath}	
\usepackage{amssymb}	

\title[SNe Ia and PPNe polarization properties]{Common continuum polarization properties: a possible link between proto-planetary nebulae and Type Ia Supernovae progenitors}

\author[A. Cikota et al.]{Aleksandar Cikota,$^{1}$\thanks{E-mail: acikota@eso.org}
Ferdinando Patat,$^{1}$
Stefan Cikota,$^{2,3}$
Jason Spyromilio,$^{1}$
\newauthor Gioia Rau$^{1,4}$
\\
$^{1}$European Southern Observatory, Karl-Schwarzschild-Str. 2, 85748 Garching b. M\"{u}nchen, Germany\\
$^{2}$University of Zagreb, Faculty of Electrical Engineering and Computing, Department of Applied Physics, Unska 3, 10000 Zagreb, Croatia\\
$^{3}$Ru{\dj}er Bo\v{s}kovi\'{c} Institute, Bijeni\v{c}ka cesta 54, 10000 Zagreb, Croatia \\
$^{4}$University of Vienna, Department of Astrophysics, T\"{u}rkenschanzstrasse 17, 1180, Vienna, Austria  \\
}

\date{Accepted XXX. Received YYY; in original form ZZZ}

\pubyear{2017}

\begin{document}
\label{firstpage}
\pagerange{\pageref{firstpage}--\pageref{lastpage}}
\maketitle

\begin{abstract}
The lines-of-sight to highly reddened SNe Ia show peculiar continuum polarization curves, growing toward blue wavelengths and peaking at $\lambda_{max} \lesssim 0.4 \mu m$, like no other sight line to any normal Galactic star. We examined continuum polarization measurements of a sample of asymptotic giant branch (AGB) and post-AGB stars from the literature, finding that some PPNe have polarization curves similar to those observed along SNe Ia sight lines. Those polarization curves are produced by scattering on circumstellar dust. We discuss the similarity and the possibility that at least some SNe Ia might explode during the post-AGB phase of their binary companion. Furthermore, we speculate that the peculiar SNe Ia polarization curves might provide observational support to the core-degenerate progenitor model.
\end{abstract}




\section{Introduction}

Supernovae Ia (SNe\,Ia) are used as cosmological probes to measure the expansion of the universe. We are confident that a SN\,Ia is an exploding C/O White dwarf (WD), close to the Chandrasekhar mass limit ($\sim$ 1.38 M$_{\odot}$, \citealt{1969Ap&SS...5..180A, 1982ApJ...253..798N}). However, the evolutionary paths and specific progenitor systems are not known. The two main progenitor scenarios are the single degenerate (SD) model, in which a WD gains mass from a nondegenerate star through accretion  \citep{1973ApJ...186.1007W}, and the double-degenerate (DD) model where two WDs merge \citep{1984ApJS...54..335I}.

Since the SD and DD models imply different circumstellar environments, studies of dust properties along the lines of sight of SNe\,Ia may provide insights on the nature of the progenitor system. SNe\,Ia observations have shown that dust along lines of sight to highly reddened SNe are characterized by low total-to-selective extinction values, $R_V \lesssim 2$ (see \citealt{2016ApJ...819..152C} for a review), compared to the average $R_V$ of $\sim$ 3.1 in the Milky Way \citep{2017ApJ...838...36S}. Low $R_V$ values can be produced by an enhanced abundance of small dust grains \citep{2003ARA&A..41..241D}, or arise as a consequence of scattering \citep{2005ApJ...635L..33W, 2005MNRAS.357.1161P}.

\citet{2015A&A...577A..53P} found that highly reddened SN\,1986G, SN\,2006X, SN\,2008fp and SN\,2014J, all characterized by low $R_V$ values (deduced from spectrophotometry), also showed peculiar non-time-variable continuum polarization wavelength properties (hereafter, we will use the term polarization curve to describe these) with peak polarization values in the blue ($\lambda_{max} \lesssim 0.4 \mu m$). We expect that the wavelength of peak polarization, $\lambda_{max}$, depends on the dust grain size distribution of aligned grains \citep{1994ApJ...431..783K}. For an enhanced abundance of aligned small dust grains, $\lambda_{max}$ shifts to shorter wavelengths, while for an enhanced abundance of aligned large dust grains to longer wavelengths \citep{1975ApJ...196..261S}. 

It remains unclear whether the peculiar polarization curves that host galaxy dust exhibit along the lines of sight to SNe\,Ia is associated to the SN itself or the more general environment. Possible explanations are that the composition of interstellar dust in SNe Ia host galaxies (or at least in some of them) is different from the dust in our Galaxy, or that there is circumstellar dust with an enhanced abundance of small grains, possibly ejected from the progenitor system before the explosion.

There is evidence that circumstellar material (CSM) might contribute to the extinction of some SNe\,Ia. \citet{Patat2007Sci...317..924P} undertook high-resolution spectroscopy of the highly reddened SN\,2006X. From observed time evolution of the Na\,{\small I}\,D doublet lines, which they attribute to changes in the CSM ionization conditions induced by the variable supernova radiation field. Similarly, \citet{2009ApJ...702.1157S} observed variable sodium absorption in SN\,2007le, a low-extinction SNe\,Ia with E(B-V) $\sim$ 0.27 mag. Furthermore, \citet{2009ApJ...693..207B} measured changes in the equivalent width of the Na\,{\small I} lines in SN\,1999cl and SN\,2006X. This detection of variable Na\,{\small I}\,D features in two of the most reddened SNe suggests that the change in equivalent width (EW) with time might be associated with circumstellar or interstellar absorption \citep{2009ApJ...693..207B}. However, no such variation is detected in the highly reddened SN\,2003cg \citep{Elias-Rosa2006MNRAS.369.1880E}. 

Another possibility is that light scattering on nearby dust contributes to both, low $R_V$ values, and a different polarization curve. \citet{2015A&A...577A..53P} illustrate, assuming that the scattering and the dichroic components are parallel, that the polarization curve of SN\,2006X may have a component induced by Rayleigh scattering in addition to the standard Serkowski component (see their Fig. 6). 

We note that no time variability in $R_V$, nor in polarization (see discussion in $\S$4.4 by \citealt{2015A&A...577A..53P}) is observed in these objects. \citet{2017ApJ...834...60Y} used \textit{HST} observations to map the ISM around SN 2014J through light echoes. They observed two echo components: a diffuse ring distributed over distances of $\sim$ 100-500 pc, and a luminous arc, produced through dust scattering of different grain sizes, at a distance  of 228 $\pm$ 7 pc from the supernova. From the wavelength dependence of the scattering optical depth, they found that the arc favors small $R_V \sim 1.4$ values, which is consistent with the $R_V$ measured along the direct line of sight, while the ring is consistent with a common Milky Way $R_V \sim 3$ value. 

\citet{2017ApJ...836...13H} fit a two-component (interstellar absorption and circumstellar scattering) dust model to photometric and (spectro)polarimetric observations of SN\,1986G, SN\,2006X, SN\,2008fp and SN\,2014J, to investigate the grain size distribution and alignment functions of dust along those lines of sight. The SN\,1986G and SN\,2006X observations could be reproduced assuming an enhanced abundance of small silicate grains in the interstellar dust only, while in the case of SN\,2014J, a contribution by circumstellar (CS) dust scattering was required. In the case of SN\,2008fp, \citet{2017ApJ...836...13H} found that the alignment of small dust grains must be as efficient as that of large grains, but the existence of CS dust could not be ascertained and suggested that the enhanced abundance of small silicate grains might be produced by cloud collisions driven by the SN radiation pressure. We note that strong SN radiation may also induce efficient alignment of small grains via the radiative torque mechanism.  

\smallskip

In this work we report a striking similarity between the polarization curves observed 
in proto-planetary nebulae (PPN) and those of highly-reddened SNe\,Ia (Fig.~\ref{fig:p-l}), and we discuss the possible implications of this finding on SNe\,Ia progenitors.

In section $\S$~2 we describe the data, explain the methods and show the results, in section $\S$~3 we discuss the results, while section $\S$~4 summarizes our conclusions.

\section{Data, methods and results}

\citet{Bieging2006ApJ...639.1053B} presented linear spectropolarimetry  of 21 asymptotic giant branch (AGB) and 13 post-AGB stars (i.e. proto-planetary nebulae, PPNe), and two R CrB stars. The polarization degree of AGB stars is smaller than that of PPNe (see their Table\,1 and Figure\,1), consistent with the findings of \citet{1991AJ....101.1735J} that  polarization increases as the evolutionary state progresses from early AGBs to PPNe, while the subsequent planetary nebula (PN) stages show lower polarization degrees than PPNe. \citet{Bieging2006ApJ...639.1053B} investigated the time variability in polarization degree of AGB stars and PPNe, finding no systematic correlation between variations in polarization and the visible luminosity in AGB stars. 
Four of five PPNe, which were observed at multiple epochs, showed no variation in polarization, indicating that they have stable dust distribution evolving on timescales longer than a few years. \citet{Bieging2006ApJ...639.1053B} group their sample of 13 PPNe in two groups: 

(i) objects with a nearly constant polarization (with fractional variation within $\sim$10\%) across the observed wavelength range ($\sim$4200--8400$\AA$), and a nearly constant polarization angle ($\pm$ 10$^{\circ}$);

\begin{figure*}
\includegraphics[trim=20mm 15mm 0mm 35mm, width=18cm, clip=true]{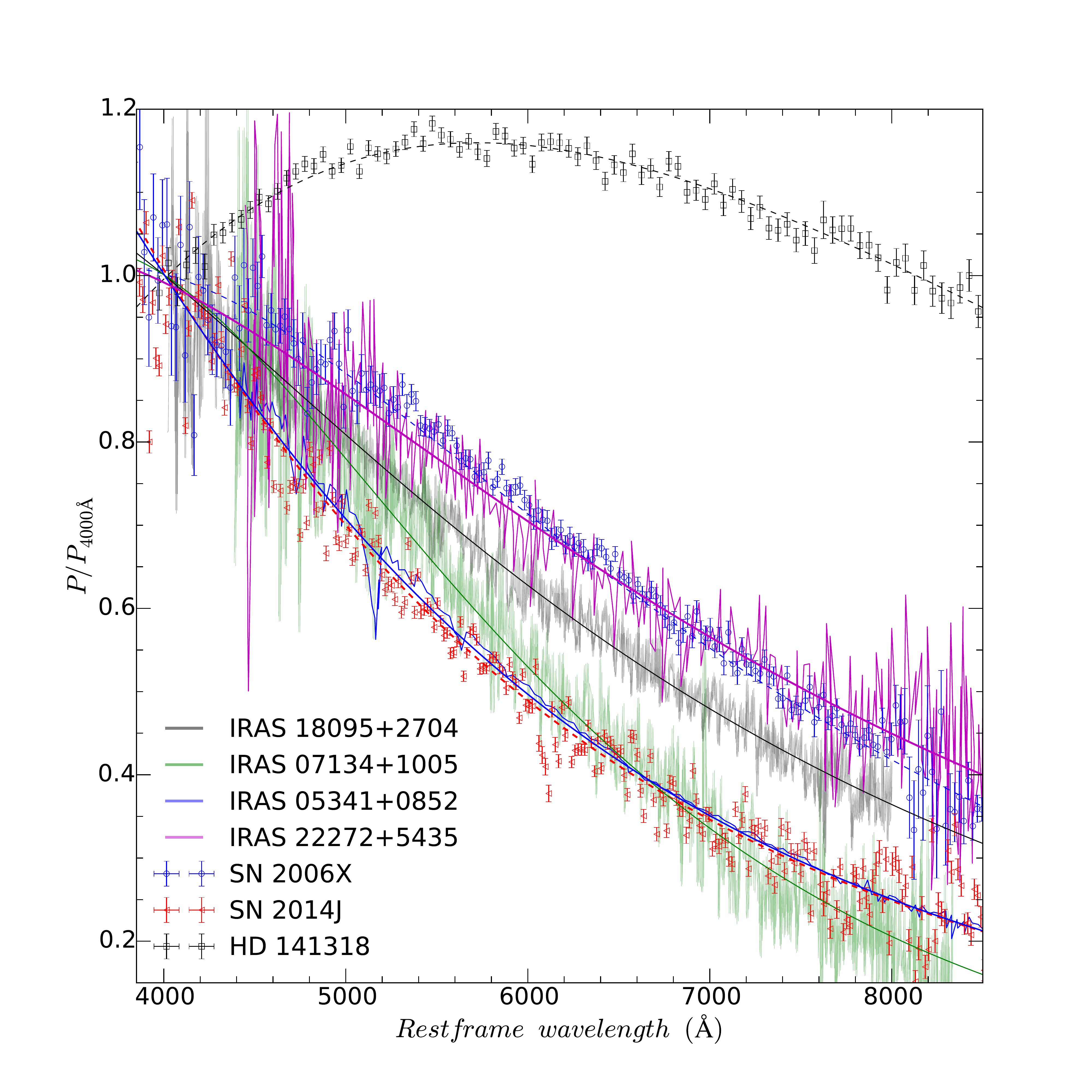}
\vspace{-3mm}
\caption{Comparison between polarization curves of SN\,2006X and SN\,2014J \citep{2015A&A...577A..53P}, HD\,141318 (Cikota et al., in prep.), and four PPNe \citep{Bieging2006ApJ...639.1053B}. HD\,141318 is a star with $R_V = 1.95 \pm 0.18$, but has a normal polarization curve with $P_{max} \sim 0.57 \mu m$, while the SNe\,Ia and PPNe have polarization curves steeply increasing towards blue wavelengths.}
\label{fig:p-l}
\end{figure*}

(ii) IRAS 05341+0852, IRAS 07134+1005, IRAS 18095+2704, and IRAS 22272+5435 -- objects with significantly rising polarization from the red ($\sim$ 8000$\AA$) to the blue ($\sim$ 4000$\AA$), while the polarization angle remains nearly constant ($\pm$ 10$^{\circ}$). Those objects have similar polarization curves to SNe\,Ia. The polarization curves are shown in Figure~\ref{fig:p-l}. \citet{Bieging2006ApJ...639.1053B} argue that those objects are similar to IRAS 04296+3429 and IRAS 08005-2356. Models by \citet{2005ApJ...624..957O} suggest that the objects present a point-symmetric morphology and evacuated lobes, respectively, and that the  increasing polarization towards the blue indicates a sharply peaked grain distribution of scattering particles. However, IRAS 04296+3429 and IRAS 08005-2356 have strongly wavelength dependent polarization angles (see Fig.\,1 in \citealt{2005ApJ...624..957O}). 

\begin{figure}
\includegraphics[trim=20mm 0mm 0mm 20mm, width=9cm, clip=true]{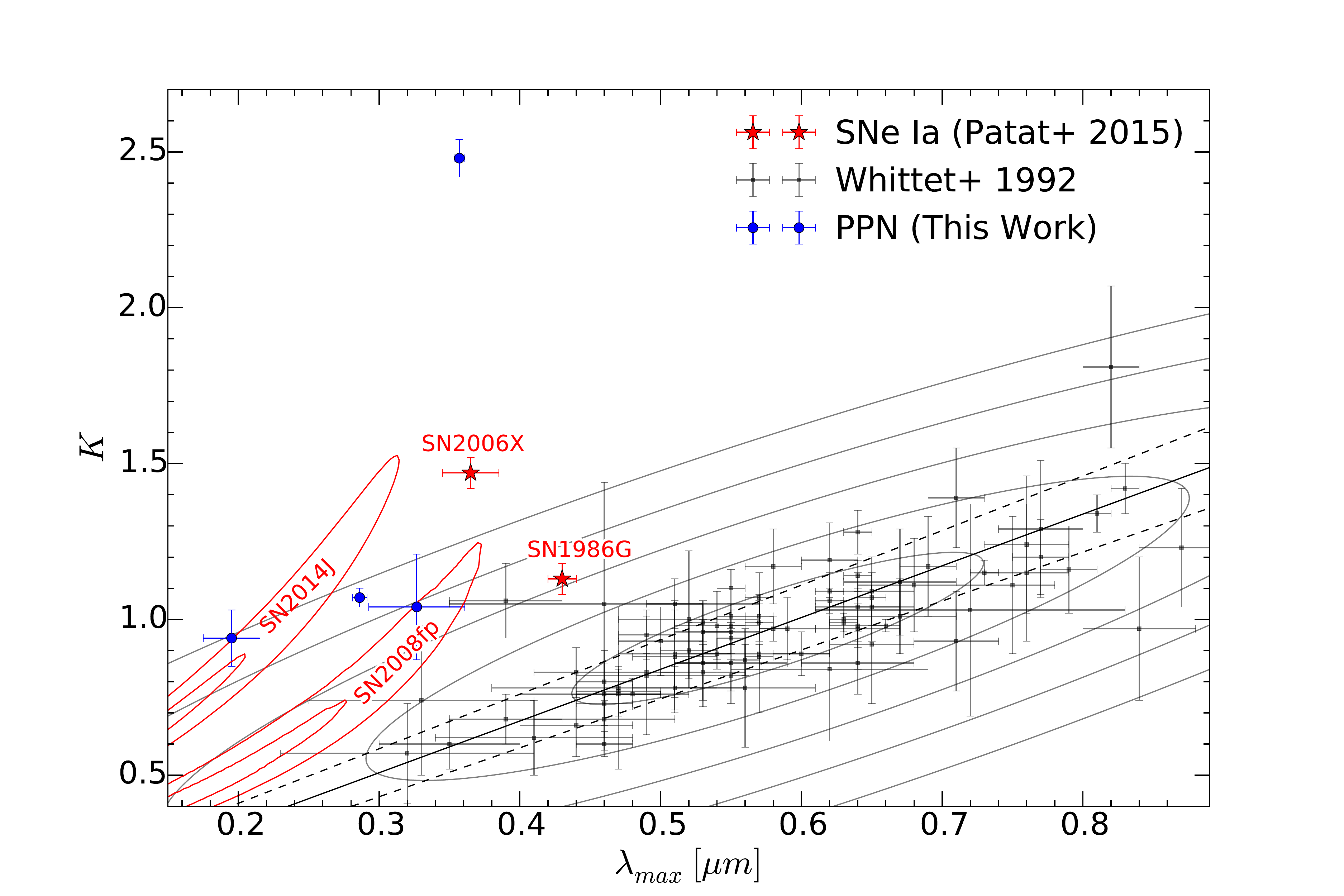}
\vspace{-5mm}
\caption{SNe\,Ia \citep{2015A&A...577A..53P} and PPNe compared to a sample of Milky Way stars \citep{1992ApJ...386..562W} in the $\lambda_{\rm max}$--K plane. The solid line is the empirical $\lambda_{\rm max}$--K relation derived by \citet{1992ApJ...386..562W}. Both, SNe\,Ia and PPNe have steeper curves (with higher K values) rising towards blue ($\lambda_{\rm max} \lesssim 0.4 \mu m$) compared to normal stars in the Milky Way.}
\label{fig:lmax_K}
\end{figure}

To parameterize the polarization curves in \citet{Bieging2006ApJ...639.1053B}, and compare them to those derived for SNe\,Ia and a sample of Galactic stars \citep{1992ApJ...386..562W}, we fit to the digitized data the empirical Serkowski curve \citep{1975ApJ...196..261S}:
\begin{equation}
\label{eq_serkowski}
\frac{P(\lambda)}{P_{max}} = \exp \left[-K ln^2 \left( \frac{\lambda_{max}}{\lambda} \right) \right] 
\end{equation}
where $\lambda_{max}$ is the wavelength at the peak polarization $P_{max}$, and $K$ depends on the width of the polarization curve (lower $K$s widen the curve). We also calculate the polarization angle, $\theta$, and its wavelength dependency through the average slope $d\theta/{d\lambda}$. The results are given in Table~\ref{tab:serkowski_table}, and shown in Fig.~\ref{fig:lmax_K}.

\begin{table*}
\centering
\caption{Serkowski parameters and polarization angles}
\label{tab:serkowski_table}
\begin{tabular}{lcccccc} 
\hline
& \multicolumn{3}{c}{Serkowski parameters} & & & \\\cline{2-4}
Name & $\lambda_{max} (\AA)$ & $P_{max}(\%)$ & $K$ & $P_{4000\AA} (\%)$& $\theta$ ($^{\circ}$) & d$\theta$/d$\lambda$ (deg $\mu$m$^{-1}$)\\
\hline
IRAS 05341+0852 & 1952 $\pm$ 202 & 25.03 $\pm$ 2.58 & 0.94 $\pm$ 0.09 & 15.43 $\pm$ 2.77 &146.3 $\pm$ 0.4 & 3.3 $\pm$ 0.5\\
IRAS 18095+2704 & 2862 $\pm$ 53  & 10.19 $\pm$ 0.12 & 1.07 $\pm$ 0.03 & 9.04 $\pm$ 0.16&15.1 $\pm$ 1.8 & -19.0 $\pm$ 1.5\\
IRAS 07134+1005 & 3570 $\pm$ 38  &  5.53 $\pm$ 0.06 & 2.48 $\pm$ 0.06 & 5.36 $\pm$ 0.07 &148.1 $\pm$ 4.8 & 32.0 $\pm$ 5.0\\
IRAS 22272+5435 & 3267 $\pm$ 341 &  4.14 $\pm$ 0.25 & 1.04 $\pm$ 0.17 & 3.97 $\pm$ 0.3&10.4 $\pm$ 1.9 & -17.5 $\pm$ 2.3\\
\hline
\end{tabular}
\end{table*}

\section{Discussion}

Fig.~\ref{fig:p-l} compares the continuum polarization curves of Type Ia SN\,2006X and SN\,2014J \citep{2015A&A...577A..53P} with those of a Galactic star with (HD~141318; $R_V = 1.95 \pm 0.18$, $P_{max} \sim 0.57 \mu m$, Cikota et al. in prep.), and four PPNe \citet{Bieging2006ApJ...639.1053B}. 

The similarity between the polarization curves observed in SNe\,Ia and PPNe is intriguing (see Fig.~\ref{fig:p-l}): both SNe\,Ia and PPNe have steep polarization curves rising towards blue wavelengths. We know of no other stellar object showing such polarization curves in our Galaxy. For PPNe we know that the bulk of polarization is produced by CS dust scattering \citep{2005ApJ...624..957O}.
This leads to the interesting possibility that also in SNe\,Ia sight lines, polarization produced by scattering of CS dust contributes to the observed continuum polarization rising towards blue wavelengths.

As discussed earlier, there may be different explanations for the polarization curves observed in highly reddened SNe Ia. They may be caused by interstellar host galaxy dust characterized by an increased abundance of small particles. Perhaps the strongest argument against our interpretation is the remarkably constant (wavelength independent) polarization angles with slopes of d$\theta$/d$\lambda$ = -0.8 $\pm$ 0.5, 1.7 $\pm$ 1.4, and -0.9 $\pm$ 1.6 deg $\mu m^{-1}$ for SN\,2006X, SN\,2008fp and SN\,2014J respectively \citep{2015A&A...577A..53P}, which would indicate that all the dust grains are aligned in the same direction due to the host galaxy magnetic field. SN\,1986G exhibited a slight wavelength dependence in the polarization angle with a slope of d$\theta$/d$\lambda$ = 4.5 $\pm$ 1.9 deg $\mu m^{-1}$ \citep{1987MNRAS.227P...1H}, that might be explained by different polarization angles of various interstellar clouds along the line of sight. \citet{2017ApJ...836...88Z} has presented a study of continuum spectropolarimetry of 19 SNe\,Ia. 8 of 12 SNe (SN\,2006X, SN\,2014J, SN\,2008fp, SN\,2010ev, SN\,2007le, SN\,2002bo, SN\,2007fb and SN\,2003W, sorted according to polarization degree, from highest) which show Na\,{\small I}\,D (a known tracer of gas, metals and dust, see e.g. Fig. 6 in \citealt{1994AJ....107.1022R}) at their host galaxy velocities, have higher polarization values with polarization curves rising towards the blue ($\lambda_{max} \lesssim 0.4 \mu m$). Moreover, their polarization angles appear to be wavelength independent (within the noise) and aligned with their host galaxy spiral arms, i.e. along the local magnetic field \citep{1996QJRAS..37..297S}.

On the other hand, the PPNe exhibit mild wavelength dependencies in their polarization angles (see Table~\ref{tab:serkowski_table}). IRAS 05341+0852 has a slope d$\theta$/d$\lambda$ = 3.3 $\pm$ 0.5 deg $\mu$m$^{-1}$, which is comparable to SN\,1986G, while IRAS 18095+2704, IRAS 07134+1005, and IRAS 22272+5435 have slopes between -19 and +32 deg $\mu$m$^{-1}$. IRAS Z02229+6208 and IRAS 22223+4327 from the first (i) group have as well constant polarization angles of 1.5 $\pm$ 1.3 and 3.9 $\pm$ 3.0 deg $\mu$m$^{-1}$ respectively (see Fig. 1 in \citealt{Bieging2006ApJ...639.1053B}), however, their polarization curves are as well nearly constant, and thus not similar to the polarization curves of SNe Ia sight lines.

The level of continuum polarization is comparably high for both PPNe and SNe Ia. The polarization degree at 4000$\AA$ is 7.8 $\pm$ 0.2 $\%$, 6.6 $\pm$ 0.1 $\%$, 5.1 $\pm$ 0.1 $\%$, and 2.6 $\pm$ 0.1 $\%$ for SN 2006X, SN 1986G, SN 2014J and SN 2008fp, respectively \citep[Table 1 in][]{2015A&A...577A..53P}, and 15.4 $\pm$ 2.8 $\%$, 9.0 $\pm$ 0.2 $\%$, 5.4 $\pm$ 0.1 $\%$, and 4.0 $\pm$ 0.3 $\%$ for IRAS 05341+0852, IRAS 18095+2704, IRAS 07134+1005, and IRAS 22272+5435, respectively (Table~\ref{tab:serkowski_table}). \citet{1991AJ....101.1735J} found that the observed \textit{UBVri} polarization degree depends of the evolutionary stage of the objects: the maximum polarization reaches $3 \%$ for red giants, $7 \%$ for late-AGB stars and $40 \%$ for PPNe, while young PNe show maximum polarizations up to $6 \%$, and true PNe only up to $\lesssim 3 \%$. \citet{1991AJ....101.1735J}  explain the differences seen in various evolutionary stages as a consequence of the circumstellar dust shell becoming more dense and aspherical as the star evolves towards a PPN, and dust dissipates after the stage of a PPN towards a PN.

All four PPNe from second group, IRAS 05341+0852, IRAS 18095+2704, IRAS 07134+1005, and IRAS 22272+5435 have been observed as part of a \textit{Hubble Space Telescope (HST)} survey of PPN candidates \citep{2000ApJ...528..861U}. They show a bright central star embedded in a faint, optically thin, elliptically elongated shell (SOLE nebulae), in contrast to the bipolar form (DUPLEX nebula) which is optically thick . 
The global polarization angle (measured on unresolved, ground-based images) is perpendicular to the major axes of IRAS 05341+0852 and IRAS 18095+2704 (see their Fig. 1). This is fully consistent with the polarization being generated by scattering and carrying the geometrical imprint of the dust distribution in the PPN. IRAS 07134+1005, and IRAS 22272+5435 are not that well resolved and have low ellipticities, so that such a comparison was not possible.
This has an immediate consequence: in a sample of PPNe, the polarization angle is expected to be randomly distributed. On the other hand, the polarization observed in highly-reddened SNe Ia shows a clear alignment along the local magnetic field. This discrepancy remains an open issue with the association proposed here (see also \citealt{2015A&A...577A..53P} and \citealt{2017ApJ...836...13H}  for more detailed discussions).

\smallskip
The sodium lines, Na\,{\small I}\,D, detected in the reddened SNe Ia spectra \citep{2017ApJ...836...88Z}, do not all necessarily originate in the ISM of the SN host galaxies. E.g. in the case of SN\,2014J there is a number of components at different velocities \citep{2015A&A...577A..53P}. As \citet{2013ApJ...779...38P} (see their $\S$~4.6) suggest, high Na abundances might be produced in nova outbursts in the SD scenario. They suggest that the "blueshifted" Na\,{\small I} profiles \citep[see][]{2011Sci...333..856S, 2012ApJ...752..101F}, and large column densities might imply a progenitor system with an AGB star phase \citep[][$\S$~4.6]{2013ApJ...779...38P}. In fact, SN\,1986G, SN\,2006X and SN\,2008fp \citep{2015A&A...577A..53P}, as well as SN\,2002bo, SN\,2007fb, and SN\,2007le \citep{2017ApJ...836...88Z} have Na\,{\small I} blueshifted lines  \citep{2011Sci...333..856S,2012ApJ...752..101F}. For the remaining SN\,2014J, SN\,2010ev and SN\,2003W, there are no measurements in \citet{2011Sci...333..856S} nor \citet{2012ApJ...752..101F}.

In the case of SN\,2008fp and SN\,2006X there is clear evidence that most of the gas along the line of sight is in a molecular cloud with strong CN features \citep{2008A&A...485L...9C,2015A&A...577A..53P}. It is also interesting to note that the time-varying features in SN\,2006X did show velocities between 50 and 100 km s$^{-1}$ \citep{Patat2007Sci...317..924P}. The time-varying features were small in terms of EW, definitely smaller than the main saturated feature arising within the molecular cloud \citep[see Fig. 1 in][]{Patat2007Sci...317..924P}. Thus, in this case, the closest time-varying material certainly does not explain the bulk of extinction. Whether the main saturated component, with a velocity difference of 50-100 km s$^{-1}$ relative to the variable features, can be attributed to a hypothetical PPN or a more distant cloud can not be distinguished from the observations. On the other hand, SN 1986G and SN 2014J have a number of Na\,{\small I} and Ca\,{\small II} components at different velocities of comparable EW \citep[see e.g. Fig. 5 in][]{2015A&A...577A..53P}, and do not show traces of CN.

Furthermore, Na\,{\small I} lines have also been observed in PPNe with bright and asymmetric circumstellar nebulae. For instance, \citet{2016ARep...60..344K} observed splitting and asymmetry of strong absorption lines, particularly of Ba\,{\small II}  (while iron absorption lines are not split nor asymmetric), in post-AGB stars with C-rich circumstellar envelopes. V5112 Sgr, which was observed during multiple nights, also shows time-variability of the shape and positions of components of the split lines, in particular Ba\,{\small II} 4934 $\AA$ line, which shows most variability. 
V448 Lac shows variability in Ba\,{\small II} 6141 $\AA$ line. The blue component of Ba\,{\small II} coincides with the blue shift of the circumstellar Na\,{\small I}\,D lines, which indicates that Ba\,{\small II} also contains, besides a stellar component, a component that forms in the circumstellar envelope. \citet{1996A&A...310..893B} detected C$_2$, CN and Na\,{\small I}\,D absorption lines in HD 56126. They determined the velocities of individual components in Na\,{\small I}\,D1 and  Na\,{\small I}\,D2, and distinguished their origin between photospheric, circumstellar and interstellar. The velocity difference between circumstellar and photospheric components is $\sim$ 14 km s$^{-1}$ (see Table 6 in \citealt{1996A&A...310..893B}). \citet{2007BaltA..16..191K} analysed spectra of HD 161796 (a post-AGB star) and measured circumstellar and interstellar Na\,{\small I}\,D lines. The CS lines are shifted by -15.9 km s$^{-1}$ relative to photospheric lines, and likely correspond to an expanding shell.

\subsection{Possible implications on the SNe\,Ia progenitor systems}

It is tempting to speculate that some SNe\,Ia with peculiar polarization curves exploded during a PPN phase of their companion star, and that the bulk of continuum polarization is produced by scattering from dust in a PPN, particularly because red giants, and late-AGB stars might play an important role in SNe\,Ia progenitor systems.

The PPN is a short phase ($\lesssim 10^4$ years) of an intermediate-mass star life, between the late AGB star and a planetary nebula (PN). The star leaves the AGB phase (i.e. the Thermal-Pulse-AGB, TP-AGB phase) once the envelope mass drops to $\sim$ 0.01 M$_{\odot}$ \citep{2001Ap&SS.275....1B, 2003ARA&A..41..391V}. During the TP-phase the star may loose more than half of its initial mass ($\sim$43 $\%$ for an initial mass of 1 M$_{\odot}$; $\sim$ 79$\%$ for an initial mass of 4 M$_{\odot}$; \citealt{2013MNRAS.434..488M, 2017ApJ...835...77M}). The velocity of the circumstellar material (CSM) ejected during the AGB phase is slow ($\gtrsim$ 10 km s$^{-1}$, see e.g. \citealt{2012Natur.490..232M}). Given an AGB lifetime of 1 Myr, and a ejecta velocity of 10 km s$^{-1}$, the material can spread to distances of up to 10 pc \citep[see also][]{2015A&A...583A.106R, 2017A&A...600A..92R}. 

It is thought that during the PPN phase, collimated fast winds (>150 km s$^{-1}$) form evacuated lobes \citep{2005ApJ...624..957O,2000ApJ...528..861U} in the previously ejected material. The PPN phase ends after the cool ($T_{eff} \lesssim 10^4 K$) post-AGB stellar core heats up to $\sim 3 \times 10^4 K$ and starts ionizing the surrounding material, becoming a PN. 

However, the majority of PNe display morphologies that can not be explained in the context of a single-star scenario. Detailed \textit{HST} observations of PNe revealed aspherical morphologies that can not be accounted for in a simple interacting stellar wind model. However, close binary stars might play a key role in the formation of aspherical PPNe and PNe and in shaping axisymmetric structures \citep{2017NatAs...1E.117J}. If the binary companion is sufficiently close, the AGB star will overflow its Roche lobe which leads to a poorly understood common envelope phase, and spiral-in of the companions orbit. Eventually the binary companions will merge \citep{2017NatAs...1E.117J, 2013A&ARv..21...59I}.
\citet{2006MNRAS.370.2004N} found that a common envelope evolution can lead to the ejection of envelope material predominantly in the orbital plane of the system, which can later be shaped by winds into bipolar structures. 

We do not observe any polarization time-variability in SNe\,Ia. Thus, the scattering material is at least at a distance of $\sim$ 350 AU (assuming an ejecta velocity of 10$^4$ km s$^{-1}$, and time of two months).
Photoionization modeling suggests that CSM must be at a distance larger than $\sim$10$^{16}$ cm from the SN explosion not to ionize Ca\,{\small II}, and cause Ca\,{\small II} time variability \citep{2009ApJ...702.1157S}. Assuming a wind velocity of 100 km s$^{-1}$, it takes only $\sim$32 years to form a cavity of that size. Also, since the SNe Ia light curve can be well approximated by a discrete duration flash time variability is expected, because at different times the light echo moves through different regions of the CSM \citep[see][]{2005ApJ...635L..33W, 2005MNRAS.357.1161P, 2006MNRAS.369.1949P}. In case of close (e.g. 0.01 ly) CSM, time variability is expected at all phases, and in case of distant (e.g. 1 ly) CSM, at late phases.

\citet{2011MNRAS.417.1466K} proposed a "core-degenerate" scenario that might explode as a Type Ia SN. They suggest that due to interaction of a WD -- post-AGB-core system with a circumbinary disc (which gets formed during the final stages of the common envelope), the orbital separation might be much more reduced than expected due to the ejection of the common envelope alone. The small separation might lead to a merger at the end of the CE phase or short time after, while the core is still hot. After the merger, the rapidly rotating WD can collapse and explode after it looses sufficient angular momentum, which can last as long as $\sim 10^6$ years \citep{2005A&A...435..967Y}. This might explain why we don't observe high continuum polarization in all SNe\,Ia.
On the other hand, \citet{2011MNRAS.417.1466K} argue that in case of nearly equal masses of two WDs, the explosion can occur during the common envelope phase \citep{2010ApJ...722L.157V}.
One should also consider the violent-merger scenario between two White Dwarfs (i.e. an AGB-core and the companion WD). It is interesting to note that one fifth of all known close-binary central stars of planetary nebulae are double degenerate \citep{2017NatAs...1E.117J}. However, the explosion mechanism should be further investigated, and is beyond the scope of this paper.

\section{Summary and conclusions}

We examined spectropolarimetric data for a sample of 21 AGB and 13 post-AGB (i.e. PPNe) stars presented in \citet{Bieging2006ApJ...639.1053B}. Our main results can be summarized as follows:

\begin{enumerate}
\item We found a remarkable similarity in polarization between a group of four PPNe and the continuum polarization curves observed in highly reddened Type Ia SN\,1986G, SN\,2006X, SN\,2008fp and SN\,2014J \citep{2015A&A...577A..53P} (see Fig.~\ref{fig:p-l}, and Fig.~\ref{fig:lmax_K}). They both have steep continuum polarization curves rising towards the blue, with $\lambda_{max} \lesssim 0.4 \mu m$, while the polarization angle is to a good approximation wavelength independent. 

\item The polarization curves rising towards the blue wavelengths in those PPNe are explained in terms of scattering on CS dust grains \citep{2005ApJ...624..957O}. No other sight lines to any normal Galactic stars have similar polarization curves. This opens the intriguing possibility that also in the case of SNe\,Ia scattering may be playing an important role.

\item Furthermore, the similarity between the polarization curves of PPNe and SNe\,Ia suggests that at least some SNe Ia may be enshrouded by a PPN. However, the observed alignment along the local magnetic field which characterizes the polarization angle of SNe Ia still needs to be reconciled with the random alignment expected for PPN.

\item Close binary stars might play a key role in formation and shaping of PPNe \citep{2017NatAs...1E.117J}. We speculate that some SNe\,Ia might explode during the post-AGB phase, as a result of a merger between a WD and a post-AGB core companion  \citep[core-degenerate progenitor model, ][]{2011MNRAS.417.1466K}. Thus, the steeply rising polarization curves towards blue wavelengths, along some SNe Ia sight lines, may provide observational support for the core-degenerate progenitor model.

\end{enumerate}

\section*{Acknowledgements}

We would like to thank to Thiem Hoang for helpful discussions. AC thanks Dominic
Bordelon and the ESO Library for preparing the BibTex file.
HD~141318 was observed with ESO Telescopes at the Paranal Observatory under Programme ID 094.C-0686.




\bibliographystyle{mnras}
\bibliography{references.bib} 







\bsp	
\label{lastpage}
\end{document}